\documentclass{INTERSPEECH2023}

\interspeechcameraready

\title{MyVoice: Arabic Speech Resource Collaboration Platform}
\name{Yousseif Elshahawy, Yassine El Kheir,  Shammur Absar Chowdhury, Ahmed Ali}
\address{
  Qatar Computing Research Institute, HBKU, Doha, Qatar}
\email{info@arabicspeech.org}
\begin{document}

\maketitle

\begin{abstract}
% ## allows admin to master the tasks 
% ## annotating
% ## Anyone can annotate

% Shammur v2
We introduce MyVoice, a crowdsourcing platform designed to collect Arabic speech to enhance dialectal speech technologies. This platform offers an opportunity to design large dialectal speech datasets; and makes them publicly available. MyVoice allows contributors to select city/country-level fine-grained dialect and record the displayed utterances. Users can switch roles between contributors and annotators. The platform incorporates a quality assurance system that filters out low-quality and spurious recordings before sending them for validation. During the validation phase, contributors can assess the quality of recordings, annotate them, and provide feedback which is then reviewed by administrators. Furthermore, the platform offers flexibility to admin roles to add new data or tasks beyond dialectal speech and word collection, which are displayed to contributors. Thus, enabling collaborative efforts in gathering diverse and large Arabic speech data. %, contributing to the advancement of dialectal speech resources. %technologies.

\end{abstract}
% \begin{abstract}
% % 1000 characters. ASCII characters only. No citations.
% Manuscripts submitted to INTERSPEECH 2023 must use this document as both an instruction set and as a template. Do not use a past paper as a template. Always start from a fresh copy, and read it all before replacing the content with your own.

% Before submitting, check that your manuscript conforms to this template. If it does not, it may be rejected. Do not be tempted to adjust the format! Instead, edit your content to fit the allowed space. The maximum number of manuscript pages is 5. The 5th page is reserved exclusively for references, which may begin on an earlier page if there is space.

% The abstract is limited to 1000 characters. The one in your manuscript and the one entered in the submission form must be identical. Avoid non-ASCII characters, symbols, maths, italics, etc as they may not display correctly in the abstract book. Do not use citations in the abstract: the abstract booklet will not include a bibliography.  Index terms appear immediately below the abstract. 
% \end{abstract}
\noindent\textbf{Index Terms}: data collection, multi-dialect Arabic, speech recognition

\section{Introduction}
% Arabic is a language with numerous dialects that vary widely from one region to another. Collecting such speech data from all these dialects in a consistent way with high quality is a challenging task. Traditional methods of collecting speech data have been limited in scope and have not been able to capture the nuances of dialectal Arabic. We believe that speech technology, like all technology, should be open and decentralized, and the \textbf{Myvoice} platform strives to achieve this goal for the Arabic world. \textbf{Myvoice} is an open-source platform that allows for the collection and sharing of high-quality speech data from a diverse range of Arabic dialects. The platform employs a permissive licensing scheme, which ensures that the data is accessible to anyone who wants to use it for research purposes.
% The availability of high-quality open-sourced speech data is particularly important for under-resourced languages, such as Arabic dialects, where there is a lack of resources for speech processing and recognition. By providing valuable utilities for handling these dialects, \textbf{Myvoice} is helping to advance the state-of-the-art in speech recognition. The remainder of the paper is organized as follows: In Section (2) we showcase the Platform Architecture, and then In Section (3) we investigate more the automatic verification system of recorded audio, then In Section (4) ...

The field of speech and language processing has been transformed by the accessibility of large datasets, empowering the creation of advanced models that exhibit outstanding performance. %However, the task of collecting extensive and diverse datasets is not devoid of challenges. 
However, the data preparation process can be costly, time-consuming, and, most importantly, encounter issues due to the under-representation of the language. These challenges can also impede progress and result in a centralized advancement, limiting the accessibility and utilization of these resources.

MyVoice\footnote{https://myvoice.arabicspeech.org/} aims to foster a collaborative community by building valuable resources that can further accelerate speech and language technology advancements while promoting open access to diverse and large datasets for everyone. The platform is designed to collect Arabic data to improve dialectal speech technologies and bridge the gap within the Arab world \cite{acm2021}, with data being accessible to everyone. %Moreover, the collective initiative will ensure the data is accessible to anyone who wants to use it for research purposes. 
MyVoice enables admins to host multiple tasks while allowing contributors to select and contribute to the tasks of their choice. For each task, a collaborator can record the displayed utterances, also can validate others recordings. The platform integrates state-of-the-art voice activity detection and dialectal speech recognition system for quality assurance and provides different statistics to the contributor and administration.

\begin{figure}[!ht]
\centering
%\vspace{-0.3cm}
\includegraphics[width=0.7\linewidth]{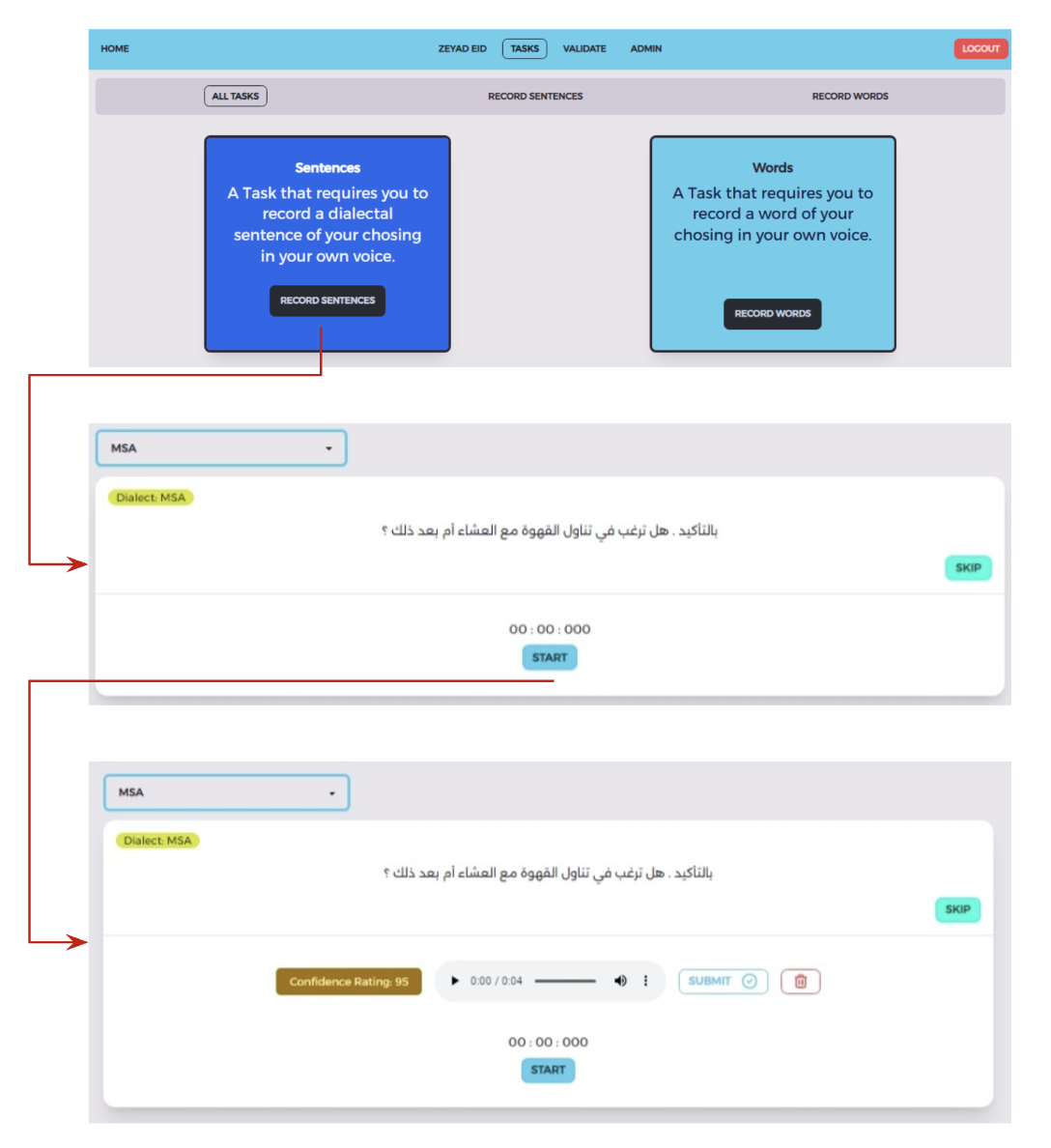}
%\vspace{-0.4cm}
\caption{\textbf{MyVoice} Tasks Page}
\label{fig:APP}
%\vspace{-0.5cm}
\end{figure}
\section{MyVoice}
% \section{MyVoice Architecture}

% \begin{figure*} [!ht]
% \centering
% \includegraphics[width=1\textwidth]
% {figures/overall_architecture.png}
% \caption{\textbf{MyVoice} Platform Functionalities}
% \label{fig:APP}
% \end{figure*}
\paragraph*{Platform Architecture:} The platform consists of two main components: a front-end and a back-end. The front-end is responsible for providing an intuitive and user-friendly interface for recording and submitting audio segments, while the back-end handles the processing, storage, and management of the submitted audio data.

The front-end is built using the following: (\textit{i})  \textbf{Nuxt3}\footnote{https://nuxt.com/}: which is a progressive framework for building web applications. It provides a powerful development experience with features such as automatic code splitting, server-side rendering, and static site generation; %\textbf{Pinia}\footnote{https://pinia.vuejs.org/}: is a lightweight and intuitive state management system for Vue.js applications. It provides a simple and reactive way to manage application state. \textbf{Tanstack Query}\footnote{https://tanstack.com/query/latest}: is a powerful data fetching library for React and Vue.js applications. It provides a declarative and composable way to fetch and manage data from APIs and other data sources. 
and (\textit{ii}) \textbf{TailwindCSS}\footnote{https://tailwindcss.com/}: which is a styling framework that provides a set of pre-defined styles and components for building responsive and modern web interfaces.

The back-end of the MyVoice platform is built using the following technologies: (\textit{i}) \textbf{FastAPI}\footnote{https://fastapi.tiangolo.com/}: FastAPI is a modern, fast (high-performance) web framework for building APIs.The Framework natively supports asynchronous programming, making it well-suited for high-traffic and data-intensive applications; %FastAPI also includes features such as automatic data validation, API documentation generation. 
(\textit{ii}) \textbf{Uvicorn}\footnote{https://www.uvicorn.org/}: Uvicorn is a fast ASGI server implementation that is built on top of the asyncio library. It's mainly used to deploy the FastAPI server for production; %\textbf{Polars}\footnote{https://www.pola.rs/}: Polars is a fast and expressive DataFrame library for Python and Rust. It provides a powerful set of tools for data manipulation and analysis and is well-suited for handling large datasets with high performance. 
(\textit{iii}) \textbf{Supabase}\footnote{https://supabase.com/}: Supabase is an open-source Tool that provides a suite of back-end services, including database management, authentication, and storage. In the MyVoice platform, Supabase is used primarily for authentication and user metadata storage, allowing users to securely log in to the platform and store their submission history and progress; %Supabase is built on top of PostgreSQL and provides a simple and intuitive API for interacting with the database, making it a powerful tool for managing user data in a scalable and secure way. 
(\textit{iv}) \textbf{PM2}\footnote{https://pm2.keymetrics.io/}: PM2 is a general process manager that is used to handle, monitor, and deploy the FastAPI server that powers the back-end of the MyVoice platform. PM2 provides features such as automatic process restarting, log management, and load balancing, which help to ensure that the server is always running smoothly and reliably. It also allows for easy deployment of updates and new features to the server, making it a key component in the development and maintenance of the MyVoice platform.

\paragraph*{Recording Tasks Interface:}
The \textbf{Tasks} page is designed to offer contributors a range of options for recording tasks and submitting their voice data as shown in Figure \ref{fig:APP}. 
% Currently, MyVoice provides two primary tasks: Arabic city-level dialectal text recording and single-word recording in Modern Standard Arabic. 
MyVoice provides the contributors with the flexibility to choose a specific dialect based on their experience and begin recording their voices given a displayed text. 
% The Arabic dialect recording task gives contributors the flexibility to choose a specific dialect based on their experience and begin recording their voices given a displayed text. 
% The flexibility of this task enables contributors to choose the dialect that they feel most comfortable with, making it easier for them to participate in the project. 
% The plaform also offers a task for recording single words in Modern Standard Arabic. This data is used to assess pronunciation and improve language learning tools. One of the great benefits of the \textbf{Tasks} page is its flexibility. 
% MyVoice allows the admin to add new utterances to the ongoing task or add a completly new tasks, ensuring that the platform can evolve and adapt to meet the needs of researchers and contributors alike.

% Administrators have the ability to add new tasks to the page, ensuring that the website can evolve and adapt to meet the needs of researchers and contributors alike.

\paragraph*{Validation Interface:}
\begin{figure}[!ht]
\centering
%\vspace{-0.8cm}
\includegraphics[width=0.7\linewidth]{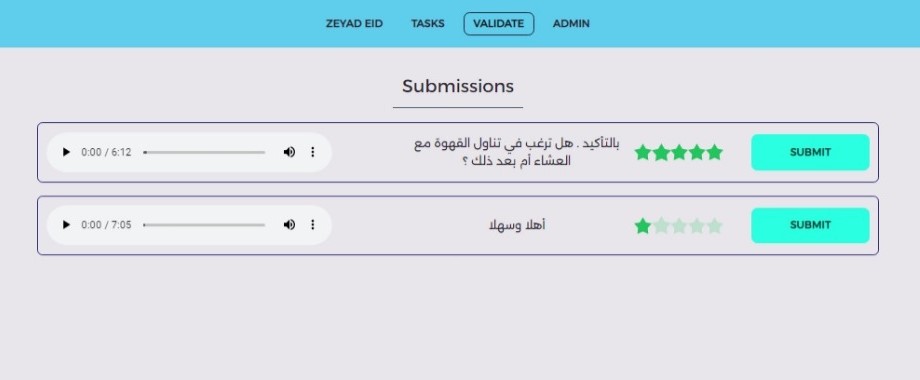}
%\vspace{-0.3cm}
\caption{\textbf{MyVoice} Audio Validation Page}
%\vspace{-0.2cm}
\label{fig:admin}
\end{figure}

The \textbf{Validation} page is a powerful tool that allows contributors to view all of their recorded audio files in one place. This page also enables contributors to assess the quality of their recordings and make decisions on whether to submit them or redo them for better quality.
By providing contributors with the ability to review their recordings and assess their quality, the \textbf{Validation} page ensures that only the highest quality recordings are submitted to the project. This additional layer of validation increases the overall quality of the voice data collected and enhances the accuracy of any research conducted using this data.

\paragraph*{Admin Interface:}
\begin{figure}[!ht]
\centering
\includegraphics[width=0.8\linewidth]{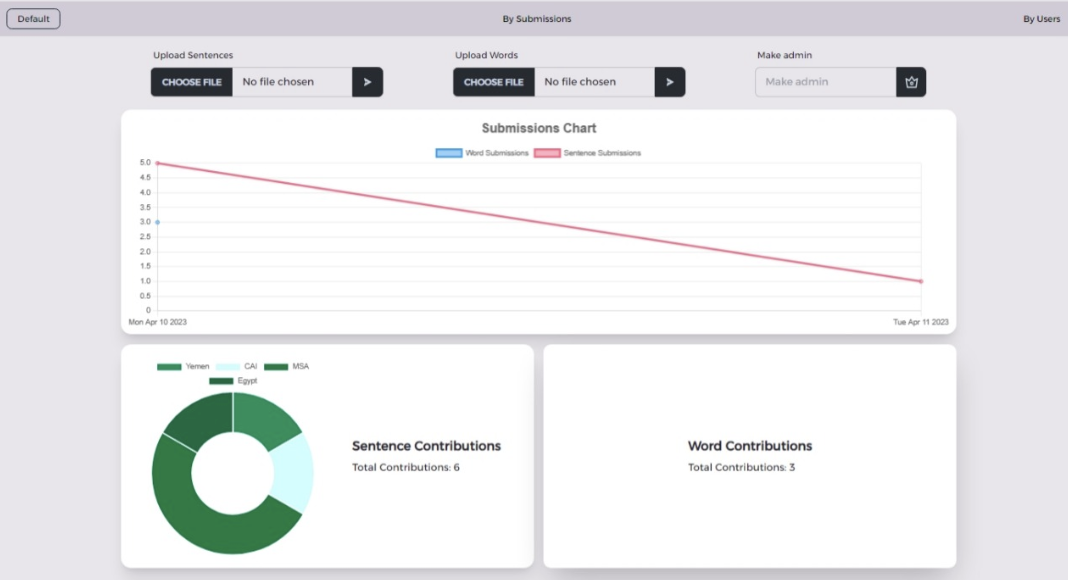}
%\vspace{-0.3cm}
\caption{\textbf{MyVoice} Admin Page}
\label{fig:admin}
%\vspace{-0.5cm}
\end{figure}
The admin page is a crucial component of the platform, it provides insights into the collected datasets and allows access to the submissions for a customizable timeline shown in Figure \ref{fig:admin}. From here, the admin can upload a new set of utterances to be recorded. The admin also has the ability to give admin privileges to any contributor.
% In addition to managing audio files, administrators can also modify the information associated with each recording. 

\begin{figure}[!ht]
\centering
\includegraphics[width=0.8\linewidth]{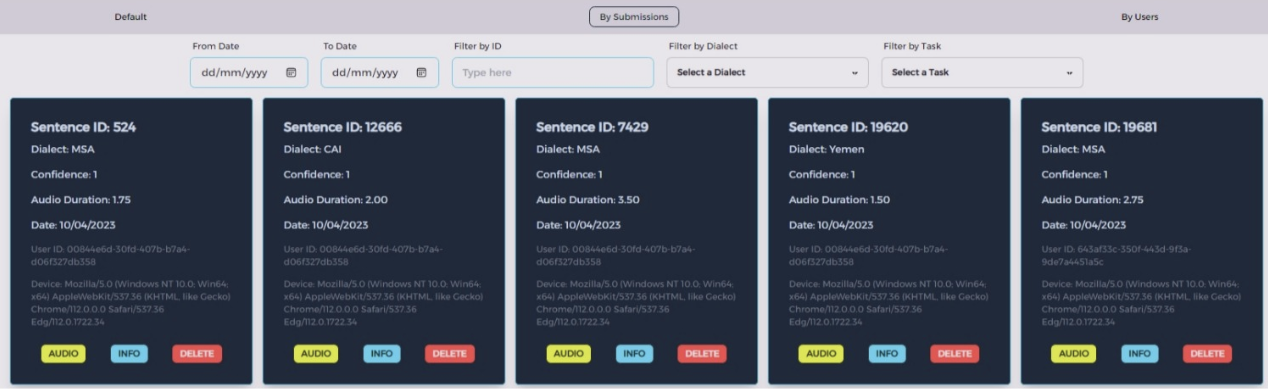}
%\vspace{-0.3cm}
\caption{\textbf{MyVoice} Submissions Page}
\label{fig:submissions}
%\vspace{-0.7cm}
\end{figure}

\paragraph*{Submissions Interface:}
The admin can also inspect individual submissions as shown in Figure \ref{fig:submissions}. Each submission is accompanied by important information, such as the task being performed and a confidence score calculated using a state-of-the-art multi-dialect Arabic speech recognition system \cite{ASR} that reflects the quality of the recording. With this information at their fingertips, administrators are able to take action based on the quality of each submission. For example, if the confidence score indicates that the recording is of poor quality or outlier recordings, administrators can delete the recordings. This ensures that all data collected by MyVoice is accurate and reliable and that contributors are held to high standards of quality.

\begin{figure}[!ht]
\centering
%\vspace{-0.3cm}
\includegraphics[width=0.8\linewidth]{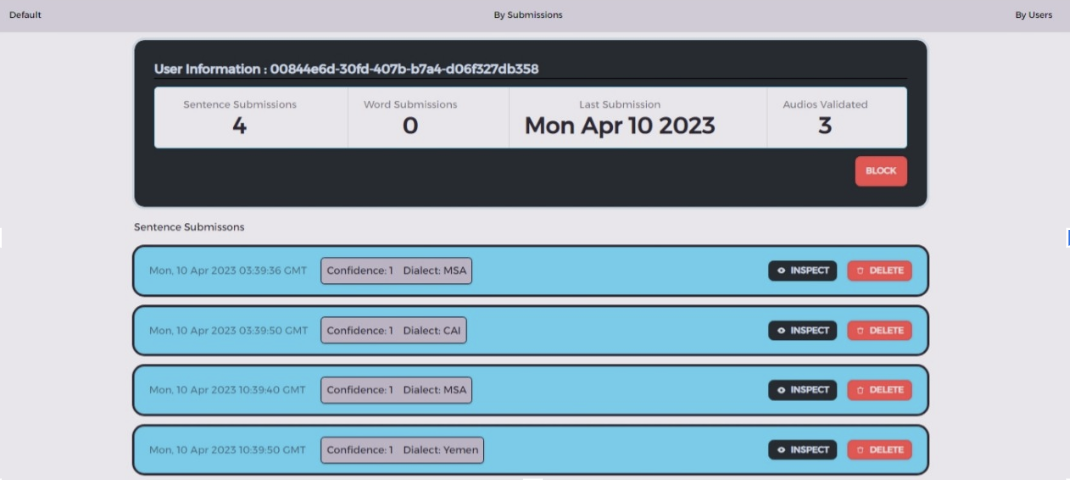}
%\vspace{-0.3cm}
\caption{\textbf{MyVoice} User Details Page}
%\vspace{-0.6cm}
\label{fig:user_details}
\end{figure}

\paragraph*{Users Interface:}
The admin can view all the current users on the page shown in Figure \ref{fig:user_details}, and from there, the admin can inspect the details of a specific user. The admin can use the details for a specific user, as shown in Figure \ref{fig:user_details}, to decide if the user is malicious or not, in which case they can block them from making further submissions as well as delete all their previous submissions.

\section{Conclusion}
% MyVoice is a powerful crowdsourcing platform that aims  advance research efforts in a variety of important areas related to Arabic dialects. By making it easier for contributors to participate and ensuring the accuracy and reliability of the voice data collected, the website is making significant contributions to fields such as speech recognition, dialect identification, and language learning for under-resourced dialects. As the website continues to evolve and grow, it will undoubtedly play an even greater role in shaping the future of research in these and other related areas.

MyVoice is a  crowdsourcing platform introduced for collecting dialectal Arabic speech to enhance dialectal speech technologies. It allows contributors to record and validate large dialectal speech datasets. The platform includes an integrated quality assurance system to aid the validator and admin to assess the recordings before making them publicly available, thus ensuring the quality of the data. It also offers statistical insights into the data and promotes collaborative efforts in gathering diverse and large dialectal Arabic speech data for further advancing speech technologies.

\bibliographystyle{IEEEtran}
\bibliography{mybib}

% Generated by IEEEtran.bst, version: 1.13 (2008/09/30)
\begin{thebibliography}{1}
\providecommand{\url}[1]{#1}
\csname url@samestyle\endcsname
\providecommand{\newblock}{\relax}
\providecommand{\bibinfo}[2]{#2}
\providecommand{\BIBentrySTDinterwordspacing}{\spaceskip=0pt\relax}
\providecommand{\BIBentryALTinterwordstretchfactor}{4}
\providecommand{\BIBentryALTinterwordspacing}{\spaceskip=\fontdimen2\font plus
\BIBentryALTinterwordstretchfactor\fontdimen3\font minus
  \fontdimen4\font\relax}
\providecommand{\BIBforeignlanguage}[2]{{%
\expandafter\ifx\csname l@#1\endcsname\relax
\typeout{** WARNING: IEEEtran.bst: No hyphenation pattern has been}%
\typeout{** loaded for the language `#1'. Using the pattern for}%
\typeout{** the default language instead.}%
\else
\language=\csname l@#1\endcsname
\fi
#2}}
\providecommand{\BIBdecl}{\relax}
\BIBdecl

\bibitem{acm2021}
A.~Ali, S.~Chowdhury, M.~Afify, W.~El-Hajj, H.~Hajj, M.~Abbas, A.~Hussein,
  N.~Ghneim, M.~Abushariah, and A.~Alqudah, ``Connecting arabs: Bridging the
  gap in dialectal speech recognition,'' \emph{Communications of the ACM},
  vol.~64, no.~4, pp. 124--129, 2021.

\bibitem{ASR}
S.~A. Chowdhury, A.~Hussein, A.~Abdelali, and A.~Ali, ``Towards one model to
  rule all: Multilingual strategy for dialectal code-switching arabic asr,''
  \emph{arXiv preprint arXiv:2105.14779}, 2021.

\end{thebibliography}
\end{document}